\documentclass[12pt]{article}
\usepackage{epsfig}
\usepackage{color}
\usepackage{amssymb,amsmath}

\newcommand{\bea}{\begin{eqnarray}}
\newcommand{\eea}{\end{eqnarray}}
\newcommand{\be}{\begin{equation}}
\newcommand{\ee}{\end{equation}}

 \makeatletter
\def\alt{\mathrel{\mathpalette\gl@align<}}
\def\agt{\mathrel{\mathpalette\gl@align>}}
\def\gl@align#1#2{\lower.6ex\vbox{\baselineskip\z@skip\lineskip\z@
\ialign{$\m@th#1\hfil##\hfil$\crcr#2\crcr\sim\crcr}}} \makeatother
\begin{document}
\begin{flushright}
\end{flushright}
\vspace*{1.0cm}

\begin{center}
\baselineskip 20pt 
{\Large\bf 
Supersymmetric minimal $B-L$ model at the TeV scale \\ 
with right-handed Majorana neutrino dark matter 
}
\vspace{1cm}

{\large 
Zachary M. Burell and Nobuchika Okada
}
\vspace{.5cm}

{\baselineskip 20pt \it
Department of Physics and Astronomy, 
University of Alabama, 
Tuscaloosa,  AL 35487, USA
} 

\vspace{.5cm}

\vspace{1.5cm} {\bf Abstract}\\
\end{center}

We propose a supersymmetric extension of the minimal $B-L$ model 
 where we consider a new $Z_2$-parity 
 under which one right-handed neutrino is assigned odd parity. 
When the Majorana Yukawa coupling of a $Z_2$-even 
 right-handed neutrino is large, radiative corrections 
 will drive the mass squared of the corresponding 
 right-handed sneutrino to negative values, 
 breaking the $B-L$ gauge symmetry at the TeV scale in a natural way. 
Additionally, R-parity is broken and thus the conventional 
 supersymmetric dark matter candidate, the neutralino, is no longer viable. 
Thanks to the $Z_2$-parity, the $Z_2$-odd right-handed neutrino 
 remains a stable dark matter candidate even in the presence 
 of R-parity violation. 
We demonstrate that the dark matter relic abundance 
 with an enhanced annihilation cross section 
 by the $B-L$ gauge boson (Z') resonance 
 is in accord with the current observations. 
Therefore, it follows that the mass of this dark matter particle 
 is close to half of the Z' boson mass. 
If the Z' boson is discovered at the Large Hadron Collider, 
 it will give rise to novel probes of dark matter: 
The observed Z' boson mass will delineate a narrow range 
 of allowed dark matter mass. 
If the Z' boson decays to a pair of dark matter particles, 
 a precise measurement of the invisible decay width 
 can reveal the existence of the dark matter particle. 

\thispagestyle{empty}

\newpage

\addtocounter{page}{-1}
\setcounter{footnote}{0}
\baselineskip 18pt
\section{Introduction}

The minimal supersymmetric (SUSY) extension of the Standard Model (MSSM) 
 is one of the prime candidates for physics beyond the Standard Model (SM), 
 which naturally solves problems in the SM, 
 in particular, the gauge hierarchy problem. 
In addition, a candidate for the cold dark matter, 
 which is missing in the SM, is also naturally 
 incorporated in the MSSM. 
Searching for SUSY is one of the major occupations
 of the Large Hadron Collider (LHC) resources. 
The LHC is operating at unprecedented luminosities, and 
 are collecting data very rapidly. 
Discovery of physics beyond the Standard Model in the near future 
 is highly likely, and anticipated.

The MSSM can solve the gauge hierarchy problem and 
 the dark matter problem, which it is able to achieve 
 merely by virtue of it being supersymmetric. 
However it is clear that the SUSY extension 
 is not enough to solve the aforementioned problems 
 in addition to explaining neutrino phenomena. 
This is because the solar and atmospheric neutrino oscillation 
 have established non-zero neutrino masses and mixings between 
 different neutrino flavors~\cite{NuData}. 
Unlike the quark sector, the scale of neutrino masses 
 is very small and the different flavors are largely mixed. 
We have no choice but to extend the MSSM in order to incorporate 
 such neutrino masses and flavor mixings. 
The seesaw extension~\cite{seesaw} has gained much attention
 since it not only accounts for the neutrino mass but also 
 explains the smallness of the mass in a natural way. 
Corresponding to the seesaw scale (the typical scale 
 of right-handed neutrinos) being, for example, 
 from 1 TeV to $10^{14}$ GeV, the scale of the neutrino 
 Dirac mass varies from 1 MeV (the electron mass scale) 
 to 100 GeV (the top quark mass scale).

The $B-L$ (baryon number minus lepton number) 
 is an anomaly-free global symmetry in the SM
 and it can be easily gauged. 
The minimal $B-L$ model is the simplest gauged $B-L$ extension 
 of the SM~\cite{B-L}, 
 where right-handed neutrinos of three generations 
 and a Higgs field with two units of the $B-L$ charge are introduced. 
The existence of the three right-handed neutrinos 
 is crucial in canceling the gauge and gravitational anomalies. 
In this model, the mass of right-handed neutrinos arises 
 associated with the $B-L$ gauge symmetry breaking.

Although the scale of the $B-L$ gauge symmetry breaking is arbitrary 
 as long as phenomenological constraints are satisfied, 
 it is interesting to consider it at the TeV scale~\cite{B-LTeV}. 
For example, it has been recently pointed out~\cite{IOO} 
 that when the classical conformal invariance is imposed 
 on the minimal $B-L$ model, the symmetry breaking scale 
 appears to be the TeV scale naturally. 
If this is the case, all new particles in the model, 
 the Z' gauge boson, the $B-L$ Higgs boson 
 and the right-handed neutrinos appear at the TeV scale, 
 which can be discovered at the LHC~\cite{LHCBL}. 
The minimal $B-L$ model also has interesting cosmological prospects 
 such as dark matter physics~\cite{OS} and baryogenesis~\cite{OOI2}.

In this paper, we investigate supersymmetric extension 
 of the minimal $B-L$ model. 
It has been pointed out~\cite{KM} that in this model 
 the $B-L$ symmetry is radiatively broken by the interplay 
 between large Majorana Yukawa couplings of right-handed neutrinos 
 and the soft SUSY breaking masses, a mechanism that is analogous 
 to radiative electroweak symmetry breaking in the MSSM. 
The mechanism naturally places the $B-L$ symmetry breaking scale 
 at the TeV scale.

Despite this remarkable feature of the SUSY minimal $B-L$ model, 
 a more thorough analysis~\cite{fateR} indicated 
 that most of the $B-L$ symmetry breaking parameter space 
 is occupied by non-zero vacuum expectation values 
 (VEVs) from right-handed sneutrinos. 
Therefore, the most likely scenario in the SUSY minimal $B-L$ model 
 with the radiative $B-L$ symmetry breaking, is that
 R-parity is violated in the vacuum. 
This means that the lightest superpartner (LSP) neutralino, 
 which is the conventional dark matter candidate 
 in SUSY models, becomes unstable and no longer remains
 a viable dark matter candidate.
As discussed in \cite{LSPgravitino}, 
 even though R-parity is broken, 
 an unstable gravitino if it is the LSP 
 has a lifetime longer than the age of the universe  
 and can still be the dark matter candidate. 
Although this is an interesting possibility, 
 a mechanism of SUSY breaking mediations 
 providing us with the LSP gravitino is limited, 
 and we do not consider the LSP gravitino in this paper.

Recently, a cogent framework for dark matter was elucidated 
 in the context of the (non-SUSY) minimal $B-L$ model~\cite{OS}, 
 where a new $Z_2$-parity was introduced and one right-handed 
 neutrino was assigned odd $Z_2$-parity 
 while the other fields were assigned even $Z_2$. 
Calculation of the relic abundance of the $Z_2$-odd right-handed 
 neutrino showed that it could account for the observed 
 relic abundance, and therefore the dark matter in our universe. 
We mention this to emphasize that we are not introducing 
 any new particle in the current model.

In this paper, we apply the same idea to the SUSY generalization 
 of the minimal $B-L$ model with the radiative $B-L$ symmetry breaking, 
 and investigate the resulting phenomenology. 
What we discovered is that the $B-L$ gauge symmetry 
 and R-parity are both broken at the TeV scale 
 by the non-zero VEV of a $Z_2$-even right-handed sneutrino, 
 for suitable regions of parameter space. 
Even in the presence of R-parity violation, 
 the $Z_2$-parity is still exact and the stability 
 of the $Z_2$-odd right-handed neutrino is guaranteed. 
Therefore, the $Z_2$-odd right-handed neutrino appears 
 to be a natural, stable dark matter candidate. 
We calculated the relic abundance of the $Z_2$-odd right-handed 
 neutrino and found that the resultant relic abundance was 
 in agreement with observations.

This paper is organized as follows. 
In the next section, we define the SUSY minimal $B-L$ model 
 with $Z_2$-parity and introduce superpotential 
 and soft SUSY breaking terms relevant for our discussion. 
In Sec.~3, we perform a numerical analysis of the renormalization 
 group equation (RGE) evolution of the soft SUSY breaking masses
 of the right-handed sneutrinos and $B-L$ Higgs fields 
 and show that the $B-L$ gauge symmetry is radiatively broken 
 at the TeV scale. 
It will be shown that one $Z_2$-even right-handed sneutrino 
 develops a VEV and hence R-parity is also radiatively broken. 
In Sec.~4, we calculate the relic abundance of 
 the right-handed neutrino and identify the parameter region 
 consistent with the observed dark matter relic abundance. 
We also discuss phenomenological constraints 
 of the model in Sec.~5. 
The last section is devoted for conclusions and discussions.

\section{Supersymmetric minimal $B-L$ model with $Z_2$-parity}

The minimal $B-L$ extended SM is based on the gauge group 
 SU(3)$_c\times$SU(2)$_L \times $U(1)$_Y \times$U(1)$_{B-L}$ 
 with three right-handed neutrinos and one Higgs scalar field 
 with $B-L$ charge $2$ but which is a singlet 
 under the SM gauge group\footnote{
Recently, a global $B-L$ extended model 
 with a dark matter candidate has been proposed~\cite{globalB-L}, 
 which has some interesting differences to the gauged model. 
}. 
As far as the motivation to introduce three generations 
 of right-handed neutrinos ($N_i^c$) is concerned, 
 the introduction of the three generations of right-handed 
 neutrinos is in no way ad-hoc; 
On the contrary, once we gauge $B-L$, their introduction 
 is forced upon us by the requirement of the gauge and 
 gravitational anomaly cancellations. 
The SM singlet scalar works to break the U(1)$_{B-L}$ 
 gauge symmetry by its VEV and at the same time, 
 generates Majorana masses for right-handed neutrinos
 which then participate in the seesaw mechanism.

It is easy to supersymmetrize this model 
 and the particle contents are listed in Table~1\footnote{
It is possible to construct a phenomenologically viable 
 SUSY $B-L$ model without $\Phi$ and $\bar{\Phi}$~\cite{MMB-L}. 
}. 
The gauge invariant superpotential relevant for our discussion is given by 
\bea 
 W_{BL} = \sum_{i=2}^3 \sum_{j=1}^3 y_D^{ij} N^c_i L_j H_u 
         + \sum_{k=1}^3 y_k \Phi N^c_k N^c_k 
         - \mu_\Phi \bar{\Phi} \Phi, 
\label{WBL} 
\eea
 where the first term is the neutrino Dirac Yukawa coupling, 
 the second term is the Majorana Yukawa coupling for 
 the right-handed neutrinos, 
 and a SUSY mass term for the SM singlet Higgs fields 
 is given in the third term. 
Without loss of generality, we have worked in the basis 
 where the Majorana Yukawa coupling matrix is real and diagonal. 
Note that Dirac Yukawa couplings between $N^c_1$ and $L_j$ 
 are forbidden by the $Z_2$-parity, so that the lightest 
 component field in $N^c_1$ is stable, 
 as long as the $Z_2$-parity is exact. 

\begin{table}[t]
\begin{center}
\begin{tabular}{c|ccc|c|c|c}
chiral superfield & SU(3)$_c$ & SU(2)$_L$ & U(1)$_Y$ 
 & U(1)$_{B-L}$ & R-parity & $Z_2$ \\
\hline
$ Q^i   $  & {\bf 3}     & {\bf 2} & $+1/6$ & $+1/3$ & $-$ & $+$ \\ 
$ U^c_i $  & {\bf 3}$^*$ & {\bf 1} & $-2/3$ & $-1/3$ & $-$ & $+$ \\ 
$ D^c_i $  & {\bf 3}$^*$ & {\bf 1} & $+1/3$ & $-1/3$ & $-$ & $+$ \\ 
\hline
$ L_i   $    & {\bf 1} & {\bf 2}& $-1/2$ & $-1$   & $-$ & $+$ \\ 
$ N^c_1   $  & {\bf 1} & {\bf 1}& $  0 $ & $+1$   & $-$ & $-$ \\ 
$ N^c_{2,3} $  & {\bf 1} & {\bf 1}& $  0 $ & $+1$ & $-$ & $+$ \\ 
$ E^c_i $  & {\bf 1} & {\bf 1}& $ -1 $ & $+1$     & $-$ & $+$ \\ 
\hline 
$ H_u$     & {\bf 1} & {\bf 2}   & $+1/2$ &  $ 0$ & $+$ & $+$ \\ 
$ H_d$     & {\bf 1} & {\bf 2}   & $-1/2$ &  $ 0$ & $+$ & $+$ \\  
$ \Phi$    & {\bf 1} & {\bf 1}   & $ 0$   &  $-2$ & $+$ & $+$ \\  
$ \bar{\Phi}$ & {\bf 1} & {\bf 1}& $ 0$   &  $+2$ & $+$ & $+$ \\  
\end{tabular}
\end{center}
\caption{
Particle contents:  
In addition to the MSSM particles, 
 three right-handed neutrino superfields ($N^c_{1,2,3}$) 
 and two Higgs superfields ($\bar{\Phi}$ and $\Phi$) 
 are introduced. 
The $Z_2$-parity for $N^c_1$ is assigned to be odd.
$i=1,2,3$ is the generation index. 
}
\end{table}

As we will discuss in the next section, 
 the $B-L$ gauge symmetry is radiatively broken at the TeV scale, 
 and the right-handed neutrinos obtain TeV-scale Majorana masses. 
The seesaw mechanism\footnote{
As we will see in the next section, R-parity is also radiatively broken. 
In this case, the right-handed neutrinos mix with the $B-L$ gaugino 
 and fermionic components of $\bar{\Phi}$ and $\Phi$,  
 and the seesaw formula is quite involved.} 
 sets the mass scale of light neutrinos at 
 $m_\nu = {\cal O}(y_D^2 v_u^2/M_R)$, 
 where $v_u$ is the VEV of the up-type Higgs doublet in the MSSM, 
 and $M_R={\cal O}$(1 TeV) is the mass scale of 
 the right-handed neutrinos. 
It is natural to assume that the mass of the heaviest light neutrino is 
 $m_\nu \sim \sqrt{\Delta m_{23}^2} \sim 0.05$ eV 
 with $\Delta m_{23}^2 \simeq 2.43 \times 10^{-3}$ eV$^2$ 
 being the atmospheric neutrino oscillation data~\cite{NuData}. 
Thus, we estimate $y_D \sim 10^{-6}$, and point out that
 such a small neutrino Dirac Yukawa coupling is negligible 
 in the analysis of RGEs.

Next, we introduce soft SUSY breaking terms 
 for the fields in the $B-L$ sector: 
\bea 
  {\cal L}_{\rm soft}&=& 
- \left( \frac{1}{2} M_{BL} \lambda_{BL} \lambda_{BL} + h.c.  \right)
- \left( \sum_{k=1}^3 m_{\tilde{N}^c_k}^2 |\tilde{N^c_k}|^2 
+ m_\Phi^2 |\Phi|^2 + m_{\bar \Phi}^2 |\bar{\Phi}|^2
  \right) 
\nonumber \\ 
&+& \left( B_\Phi \bar{\Phi} \Phi 
 + \sum_{k=1}^3 A_k \Phi \tilde{N^c_k} \tilde{N^c_k} + h.c. \right).
\eea
Here we have omitted terms relevant to the neutrino Dirac Yukawa couplings 
 since they are very small, i.e.  $O(10^{-6})$ or smaller. 
For simplicity, in this analysis we consider
 the same setup as the constrained MSSM 
 and assume the universal soft SUSY breaking parameters, 
 $m_{\tilde{N}^c_k}^2 = m_\Phi^2 = m_{\bar \Phi}^2 = m_0^2$ 
 and $A_k = A_0$, 
 at the grand unification scale\footnote{
However, we do not necessarily assume grand unification 
 behind our model.  
In fact, it is very non-trivial to unify the $Z_2$-odd 
 right-handed neutrino with $Z_2$-even fields. 
}, $M_U =2 \times 10^{16}$ GeV.

Before closing this section, we comment on the uniqueness 
 of the $Z_2$-parity assignment in the phenomenological 
 point of view. 
One may find the $Z_2$-parity assignment ad-hoc, 
 but we cannot assign an odd-parity for any MSSM particles 
 because the parity forbids the Dirac Yukawa couplings 
 which is necessary to reproduce the observed fermion masses 
 and quark flavor mixings. 
As we will see in the next section, the scalars 
 $\Phi$ and $\bar{\Phi}$ develop non-zero VEVs 
 to break the $B-L$ gauge symmetry, and 
 these fields should be $Z_2$-parity even 
 in order to keep the parity unbroken. 
Hence, we can assign $Z_2$-odd parity only 
 for right-handed neutrinos. 
Considering the fact that we need at least two right-handed neutrinos 
 to reproduce the observed neutrino oscillation data, 
 two right-handed neutrinos should be parity even 
 and be involved in the seesaw mechanism. 
As a result, we have assigned $Z_2$-parity odd 
 for only one right-handed neutrino as in Table~1.

\section{Radiative $B-L$ symmetry breaking and R-parity} 

In the non-SUSY minimal $B-L$ model, the $B-L$ symmetry breaking 
 scale is determined by parameters in the Higgs potential 
 which can in general be at any scale as long as 
 the experimental constraints are satisfied. 
The LEP experiment has set the lower bound 
 on the $B-L$ symmetry breaking scale as 
 $m_{Z'}/g_{BL} \geq 6-7$ TeV~\cite{vBL}. 
Recent LHC results for Z' boson search 
 with 1.1 fb$^{-1}$ \cite{LHCZ'1} 
 excluded the $B-L$ Z' gauge boson mass 
 $m_{Z'} \lesssim 1.5$ TeV \cite{Basso} 
 when the $B-L$ coupling is not too small. 
We see that the LEP bound is more severe 
 than the LHC bound for $m_{Z'} \gtrsim 1.5$ TeV. 
The SUSY extension of the model, however, 
 offers a very interesting possibility for constraining 
 the $B-L$ symmetry breaking scale, as pointed out in \cite{KM}.

It is well-known that the electroweak symmetry breaking 
 in the MSSM is triggered by radiative corrections 
 to the up-type Higgs doublet mass squared 
 via the large top Yukawa coupling~\cite{REWSB}. 
Directly analogous to this situation, the $B-L$ symmetry breaking 
 occurs through radiative corrections 
 with a large Majorana Yukawa coupling.

\begin{figure}[t]
\begin{center}
{\includegraphics[scale=1.0]{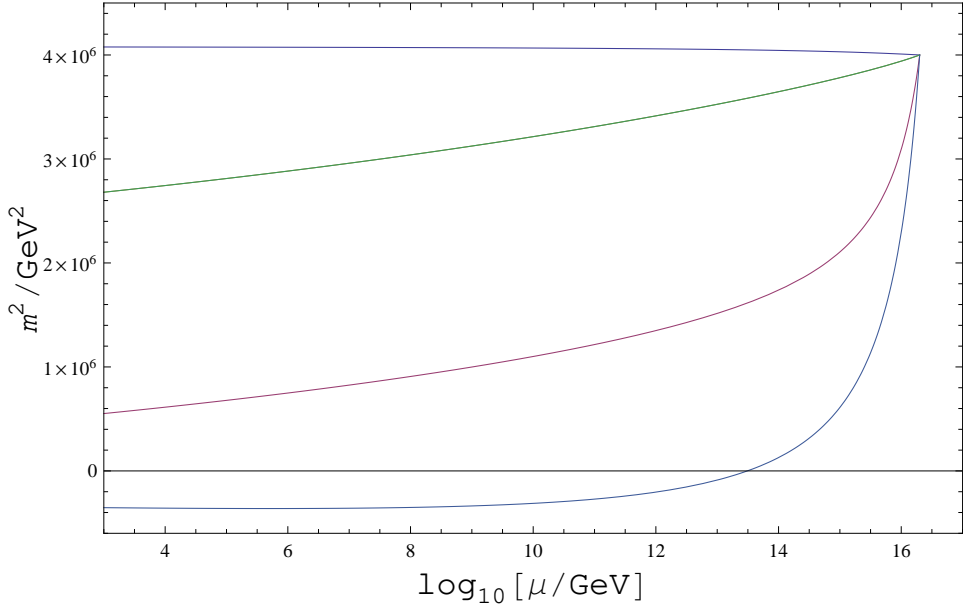}}
\caption{
The RGE running of the soft SUSY breaking masses, 
 $m_{\bar \Phi}^2$, $m_{\tilde{N}^c_1}^2 =m_{\tilde{N}^c_2}^2$, 
 $ m_{\Phi}^2$ and $m_{\tilde{N}^c_3}^2$ 
 from above. 
}
\label{fig1}
\end{center}
\end{figure}

We consider the following RGEs for soft SUSY breaking terms 
 in the $B-L$ sector~\cite{fateR, RGE2} :  
\bea 
 16 \pi^2 \mu \frac{d M_{BL}}{d \mu} 
&=&  48 g_{BL}^2 M_{BL}, \nonumber \nonumber \\ 
 16 \pi^2 \mu \frac{d m_{\tilde{N}^c_i}^2}{d \mu} 
&=&   8 y_i^2 m_\Phi^2 + 16 y_i^2 m_{\tilde{N}^c_i}^2
  + 8 A_i^2 - 8 g_{BL}^2 M_{BL}^2,  \nonumber \\
 16 \pi^2 \mu \frac{d m_{\Phi}^2}{d \mu} 
&=&   4 \left( \sum_{i=1}^3 y_i^2 \right) m_{\Phi}^2
 + 8 \sum_{i=1}^3 y_i^2 m_{\tilde{N}^c_i}^2 
 + 4 \sum_{i=1}^3 A_i^2 -32 g_{BL}^2 M_{BL}^2, \nonumber \\ 
 16 \pi^2 \mu \frac{d m_{\bar \Phi}^2}{d \mu} 
&=&  -32 g_{BL}^2 M_{BL}^2, \nonumber \\ 
 16 \pi^2 \mu \frac{d A_i}{d \mu}  
&=&  
\left(
  30 y_i^2 + 2 \sum_{j \neq i} y_j^2 - 12 g_{BL}^2  \right) A_i
+ 4 y_i \left( \sum_{j \neq i} y_j A_j - 6 g_{BL}^2 M_{BL} 
 \right),  
\eea  
 where RGEs for the gauge and Yukawa couplings are given by 
\bea 
  16 \pi^2 \mu \frac{d g_{BL}}{d \mu} 
&=&  24 g_{BL}^3 , \nonumber \\ 
  16 \pi^2 \mu \frac{d y_i}{d \mu}   
&=& y_i 
  \left( 
  10 y_i^2 + 2 \sum_{j \neq i} y_j^2 -12 g_{BL}^2 
  \right). 
\eea
To illustrate the radiative B-L symmetry breaking, 
 we solve these equations from $M_U=2 \times 10^{16}$ GeV 
 to low energy, choosing the following boundary conditions. 
\bea 
&& g_{BL}= 0.532, \; \; 
   y_1 = y_2 = 0.4, \; y_3= 2.5, \nonumber \\ 
&& M_{BL}=500 \; {\rm GeV}, \; \; 
   m_{\tilde{N}^c_i} = m_{\Phi} = m_{\bar \Phi} = 2 \; {\rm TeV}, \; \; 
   A_i = 0.   
\label{BC}
\eea 
The RGE running of soft SUSY breaking masses 
 as a function of the renormalization scale 
 is shown in Fig.~\ref{fig1}. 
After the RGE running, $m_{\tilde{N}^c_3}^2$ becomes negative
 while the other squared masses remain positive. 
The negative mass squared of the right-handed sneutrino 
 triggers not only the $B-L$ symmetry breaking 
 but also R-parity violation. 
Detailed analysis with random values of parameters 
 has shown~\cite{fateR} that 
 in most of the parameter region 
 realizing the radiative $B-L$ symmetry breaking, 
 R-parity is also  broken. 

We now analyze the scalar potential with the soft SUSY breaking 
 parameters obtained from solving RGEs. 
Since the $B-L$ symmetry breaking scale is set to be 7 TeV 
 in the following, we evaluate the RGE solutions at 7 TeV as follows: 
\bea 
&& g_{BL}= 0.286, \; \; 
   y_1 = y_2 = 0.304, \; 
   y_3= 0.561, \nonumber \\ 
&& M_{BL}=144 \; {\rm GeV}, \; \; 
   m_{\tilde{N}^c_1}^2 = m_{\tilde{N}^c_2}^2 
  = 2.73 \times 10^6 \; {\rm GeV}^2, \; \; 
   m_{\tilde{N}^c_3}^2 = -3.57 \times 10^5 \; {\rm GeV}^2, 
\nonumber \\ 
&& m_{\Phi}^2 = 6.03 \times 10^5 \; {\rm GeV}^2, \; \; 
   m_{\bar \Phi}^2 = 4.08 \times 10^6 \; {\rm GeV}^2, 
\nonumber \\ 
&& A_1 = A_2 = 34.1 \; {\rm GeV}, \; \; 
   A_3 = 25.2 \; {\rm GeV}. 
\label{input}
\eea 
The scalar potential for $\tilde{N}^c_3$, $\Phi$ and $\bar{\Phi}$ 
 consists of supersymmetric terms and soft SUSY breaking terms, 
\bea 
V= V_{SUSY} +V_{Soft},  
\eea  
where 
\bea 
&& V_{SUSY}= |2 y_3 \tilde{N}^c_3 \Phi|^2 + |\mu_\Phi \Phi|^2 
        + |y_3 (\tilde{N}^c_3)^2 - \mu_\Phi \bar{\Phi}|^2 
        + \frac{g_{BL}^2}{2} 
\left(|\tilde{N}^c_3|^2 - 2 |\Phi|^2 + 2 |\bar{\Phi}|^2 \right)^2,  
\nonumber \\
&& V_{Soft} = 
  m_{\tilde{N}^c_3}^2 |\tilde{N}^c_3|^2 
 + m_{\Phi}^2 |\Phi|^2
 + m_{\bar \Phi}^2 |{\bar \Phi}|^2
 - \left( 
  A_3 \Phi \tilde{N}^c_3 \tilde{N}^c_3 + B_\Phi \bar{\Phi} \Phi + h.c. 
  \right). 
\eea
With appropriate values of $\mu_\Phi$ and $B_\Phi$, 
 it is easy to numerically solve the stationary conditions 
 for the scalar potential. 
For example, we find (in units of GeV)  
\bea 
 \langle \tilde{N}^c_3 \rangle = \frac{2762}{\sqrt 2}, \; \; 
 \langle \Phi \rangle =\frac{2108}{\sqrt 2}, \; \;
 \langle \bar{\Phi} \rangle = \frac{2429}{\sqrt 2} 
\eea
 for $\mu_\Phi=2695$ GeV, $B_\Phi= 1.019 \times 10^7$ GeV$^2$ 
 and the parameters given in Eq.~(\ref{input}). 
In this case, we have the Z' boson mass  
\bea 
  m_{Z'} = g_{BL} v_{BL} = 2 \; {\rm TeV}, 
\eea 
where 
\bea 
  v_{BL} = 
  \sqrt{
   2 \langle \tilde{N}^c_3 \rangle^2 
  + 8 \langle \Phi \rangle^2 + 8 \langle \bar{\Phi} \rangle^2} 
  = 7 \; {\rm TeV}
\eea 
 and the experimental lower bound 
 $v_{BL} \geq 6-7$ TeV~\cite{vBL} is satisfied.

In order to prove that the stationary point is actually 
 the potential minimum, we calculate the mass spectrum 
 of the scalars, $\tilde{N}^c_3$, $\Phi$ and $\bar{\Phi}$. 
By straightforward numerical calculations, 
 we find the eigenvalues of the mass matrix of the scalars 
 $\Re[\tilde{N}^c_3]$, $\Re[\Phi]$ and $\Re[\bar{\Phi}]$ as 
 $(1035, 1868, 5315)$ in GeV, 
 while the mass eigenvalues for the pseudo-scalars 
 $\Im[\tilde{N}^c_3]$, $\Im[\Phi]$ and $\Im[\bar{\Phi}]$ 
 as $(0, 3308, 4858)$ in GeV. 
As expected, there is one massless eigenstate 
 corresponding to the would-be Nambu-Goldstone mode. 
The other right-handed sneutrino mass eigenvalues are given by 
\bea  
m_{\tilde{N}_{R i}}^2  = 
 m_{\tilde{N}^c_i}^2 
 + 4 y_i^2 \langle \Phi \rangle^2 
 - 2 y_i y_3 \langle \tilde{N}^c_3 \rangle^2 
 + 2 A_i \langle \Phi \rangle 
 + 2 y_i \mu_\Phi \langle \bar{\Phi} \rangle 
 + D_{BL},  
\nonumber \\
m_{\tilde{N}_{I i}}^2  = 
 m_{\tilde{N}^c_i}^2 
 + 4 y_i^2 \langle \Phi \rangle^2 
 + 2 y_i y_3 \langle \tilde{N}^c_3 \rangle^2 
 - 2 A_i \langle \Phi \rangle 
 - 2 y_i \mu_\Phi \langle \bar{\Phi} \rangle 
 + D_{BL}, 
\eea
where $m_{\tilde{N}_{R i}}$ and $m_{\tilde{N}_{I i}}$ 
 ($i=1,2$) are the mass eigenvalues for scalars and pseudo-scalars, 
 respectively, and 
 $D_{BL}=g_{BL}^2 ( \langle \tilde{N}^c_3 \rangle^2 
-2 \langle \Phi \rangle^2  + 2 \langle \bar{\Phi} \rangle^2 ) $. 
We find $m_{\tilde{N}_{R 1}} = m_{\tilde{N}_{R 2}} = 2626$ GeV 
 and $m_{\tilde{N}_{I 1}} = m_{\tilde{N}_{I 2}} = 1035$ GeV. 
Since the fermion components in 
 $N^c_{2,3}$, $\Phi$ and $\bar{\Phi}$ and the $B-L$ gauginos 
 are all mixed, it is quite involved to find the Majorana fermion 
 mass eigenvalues. 
Accordingly, the seesaw mechanism is realized in a very complicated way. 
Although we do not discuss the fermion spectrum in detail here, 
 our system with two right-handed neutrinos coupling to 
 the SM neutrinos provides many free parameters;
 enough to reproduce the observed neutrino oscillation data. 
On the other hand, the mass of the $Z_2$-odd 
 right-handed neutrino $N^c_{1}$ is simply given by 
\bea  
 M_{N^c_1} = 2 y_1 \langle \Phi \rangle = 906 \; {\rm GeV}.
\eea

\begin{figure}[t]
\begin{tabular}{cc}
\begin{minipage}{0.5\hsize}
\begin{center}
{\includegraphics[scale=.9]{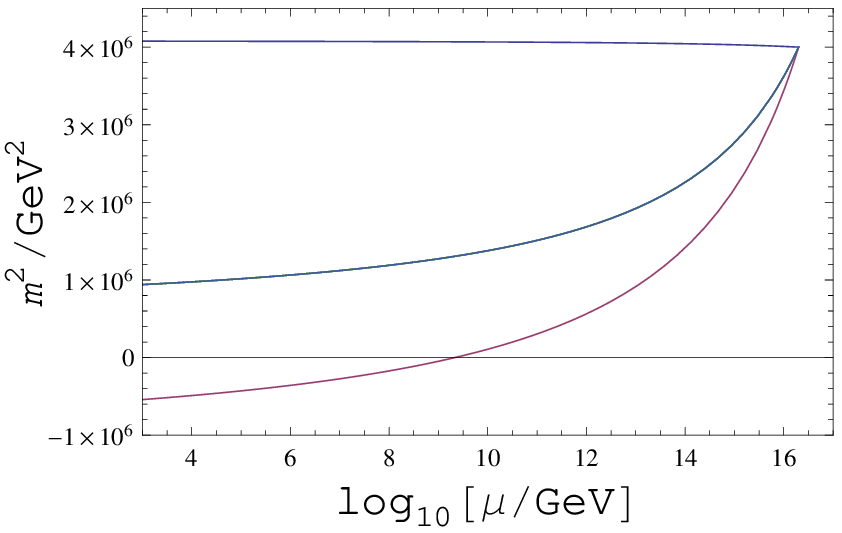}}
\end{center}
\end{minipage}
\begin{minipage}{0.5\hsize}
\begin{center}
{\includegraphics[scale=.9]{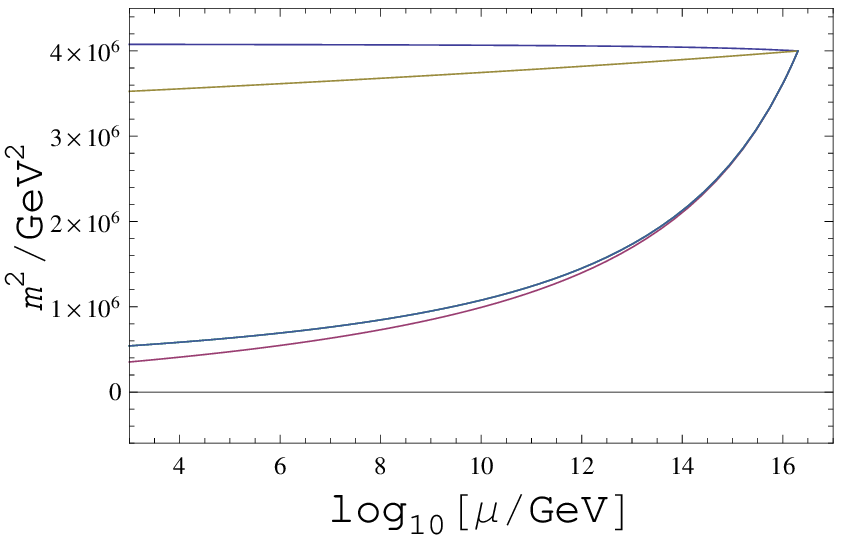}}
\end{center}
\end{minipage}
\end{tabular}
\caption{
The RGE runnings of the soft SUSY breaking masses.  
The left panel shows the results 
 for the boundary condition $y_1=y_2=y_3=1$ 
 and others are the same as in Eq.~(\ref{BC}), 
 $m_{\bar \Phi}^2$, 
 $m_{\tilde{N}^c_1}^2 =m_{\tilde{N}^c_2}^2=m_{\tilde{N}^c_3}^2$
 and $m_{\Phi}^2$ from above. 
The results for the boundary condition $y_1=0.2$, $y_2=y_3=1$ 
 are shown in the right panel, 
 where 
 $m_{\bar \Phi}^2$, $m_{\tilde{N}^c_1}^2$, 
 $m_{\tilde{N}^c_2}^2=m_{\tilde{N}^c_3}^2$  
 and $m_{\Phi}^2$ from above. 
}
\label{fig2}
\end{figure}

Before concluding this section, it is interesting  
 to observe the RGE running of the soft masses 
 for different choices of the Majorana Yukawa couplings. 
The results are depicted in Fig.~\ref{fig2}. 
In the left panel, we have taken the boundary condition 
 $y_1=y_2=y_3=1$ but other parameters are the same 
 as in Eq.~(\ref{BC}), 
 while the right panel shows the results for $y_1=0.2$, $y_2=y_3=1$. 
In general, we can categorize the results of RGE runnings 
 into three cases: 
(i) for only one $y_i ={\cal O}$(1) and the others are 
 relatively small, $m_{\tilde{N}^c_i}^2$ is driven to be negative, 
 as a result, the $B-L$ symmetry as well as R-parity is broken 
 at low energies. 
 This case corresponds to Fig.~\ref{fig1}. 
(ii) for $y_1 \sim y_2 \sim y_3 ={\cal O}$(1), 
 only $m_\Phi^2$ is driven to be negative. 
In this case, the $B-L$ symmetry is broken 
 while R-party remains unbroken. 
This case corresponds to the left panel in Fig.~\ref{fig2}. 
(iii) for $y_i \sim y_j ={\cal O}$(1) ($i \neq j$) and 
 the other $y_k$ ($k \neq i,j$) is relatively small, 
 all scalar squared masses remain positive and hence 
 no symmetry breaking occurs. 
The right panel in Fig.~\ref{fig2} corresponds to this case.

\section{Right-handed neutrino dark matter} 
As we showed in the previous section, the $B-L$ gauge symmetry 
 is radiatively broken at the TeV scale. 
Associated with this radiative breaking, 
 the right-handed sneutrino ${\tilde N}^c_3$ 
 develops  VEV and as a result, R-parity is also broken. 
Therefore, the neutralino is no longer the dark matter candidate. 
However, note that in our model the $Z_2$-parity is still exact, 
 by which the lightest $Z_2$-odd particle is stable and 
 can play the role of dark matter 
 even in the presence of R-parity violation. 
As is evident in the mass spectrum we found in the previous section, 
 the right-handed neutrino $N^c_1$ is the lightest $Z_2$-odd particle. 
In this section, we evaluate the relic abundance of 
 this right-handed neutrino dark matter candidate 
 and identify the parameter region(s) consistent 
 with the observations.

In \cite{OS}, the relic abundance of the right-handed neutrino 
 dark matter is analyzed in detail, 
 where annihilation processes through the SM Higgs boson 
 in the $s$-channel play the crucial role to reproduce 
 the observed dark matter relic abundance. 
In the non-SUSY minimal $B-L$ model, the right-handed neutrino 
 has a sizable coupling with the SM Higgs boson 
 due to the mixing between the SM Higgs doublet and 
 the $B-L$ Higgs in the scalar potential. 
However, in supersymmetric extension of the model 
 there is no mixing between the MSSM Higgs doublets 
 and the $B-L$ Higgs superfields in the starting superpotential. 
Although such a mixing emerges through the neutrino Dirac Yukawa 
 coupling with the VEV of the right-handed sneutrino 
 ${\tilde N}^c_3$, it is very small because of the small 
 neutrino Dirac Yukawa coupling $y_D={\cal O}(10^{-6})$. 
Among several annihilation channels of 
 a pair  of the $Z_2$-odd right-handed neutrinos, 
 we find that the $s$-channel Z' boson exchange process gives 
 the dominant contribution.

Now we evaluate the relic abundance of the right-handed neutrino 
 by integrating the Boltzmann equation~\cite{KT}, 
\bea 
 \frac{dY_{N^c_1}}{dx}
 =-\frac{x \gamma_{Z^\prime}}{sH(M)} 
 \left[ \left(\frac{Y_{N^c_1}}{Y_{N^c_1}^{eq}}\right)^2-1 \right],  
\label{Boltmann}
\eea  
 where $Y_{N^c_1}$ is the yield (the ratio of the number density to 
 the entropy density $s$) of the $Z_2$-odd right-handed neutrino, 
 $Y_{N^c_1}^{eq}$ is the yield in thermal equilibrium, 
 temperature of the universe is normalized by the mass 
 of the right-handed neutrino $x=M/T$, 
 and $H(M)$ is the Hubble parameter at $T=M$. 
The space-time densities of the scatterings mediated 
 by the $s$-channel Z' boson exchange in thermal equilibrium 
are given by 
\bea 
\gamma_{Z^\prime} = 
 \frac{T}{64\pi^4} \int^\infty_{4 M^2} 
 ds \hat{\sigma}(s) \sqrt{s} K_1 \left(\frac{\sqrt{s}}{T}\right), 
\eea 
where $s$ is the squared center-of-mass energy, 
 $K_1$ are the modified Bessel function of the first kind, 
 and the total reduced cross section 
 for the process $N^c_1 N^c_1 \rightarrow Z' \rightarrow f \bar{f}$
 ($f$ denotes the SM fermions) is  
\bea 
 \hat{\sigma}_{Z^\prime}(s)
 = \frac{26}{12 \pi} g_{BL}^4 
 \frac{\sqrt{s} \left( s - 4 M^2\right)^{\frac{3}{2}}}
 {\left(s-m_{Z'}^2 \right)^2 + m_{Z'}^2 \Gamma_{Z'}^2}
\eea 
 with the decay width of the Z' boson,  
\bea 
\Gamma_{Z'} = 
 \frac{g_{BL}^2}{24 \pi} 
 \left[ 13 + 2 \left( 1-\frac{4 M^2}{m_{Z'}^2} \right)^{\frac{3}{2}} 
 \theta \left( m_{Z'}^2/M^2 - 4 \right)  \right]. 
\label{width}
\eea 
For simplicity, we have assumed that $y_1=y_2$ as in the previous section 
 and that the other particles (except for the SM particles) 
 are all heavy with mass $> m_{Z'}/2$. 
This assumption is consistent with the parameter choice 
 in our analysis below.

\begin{figure}[t]
\begin{center}
{\includegraphics[scale=0.8]{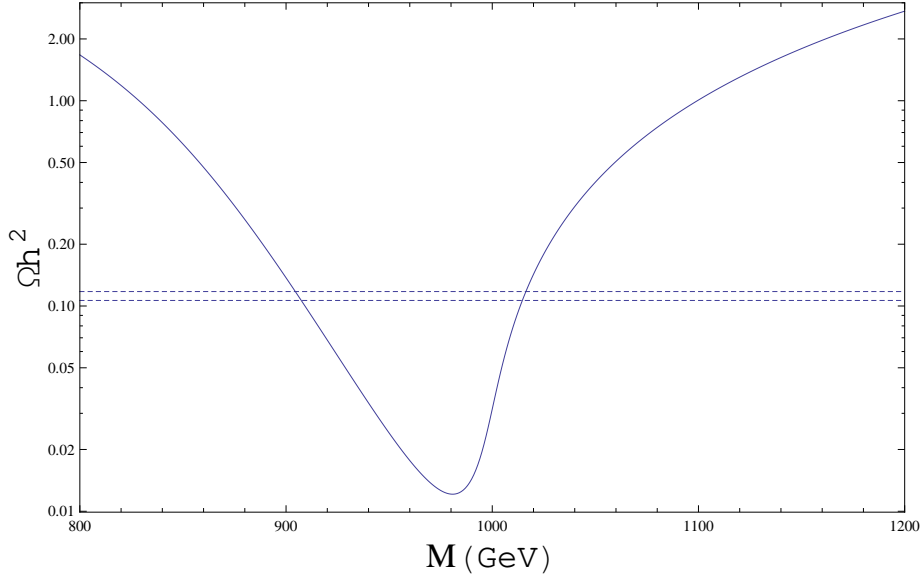}}
\caption{
The relic abundance of the dark matter right-handed neutrino 
 as a function of its mass. 
The dashed lines represent the upper and the lower bounds 
 on the dark matter relic abundance. 
}
\label{fig3}
\end{center}
\end{figure}

Now we solve the Boltzmann equation numerically. 
To solve the equation for the relevant domain, we inherit parameter values from those presented in the previous section which were already motivated as interesting values, 
\bea 
 g_{BL}=0.286, \; \; m_{Z'}=2 \; {\rm TeV}, 
\eea
 while $M_{N^c_1}=M$ is taken to be a free parameter. 
With the asymptotic value of the yield $Y_{N^c_1}(\infty)$ 
 the dark matter relic density is written as 
\bea 
  \Omega h^2 =\frac{M s_0 Y_{N^c_1}(\infty)}{\rho_c/h^2}, 
\eea 
 where $s_0 = 2890$ cm$^{-3}$ is the entropy density 
 of the present universe, 
 and $\rho_c/h^2 =1.05 \times 10^{-5}$ GeV/cm$^3$ 
 is the critical density.
The result should be compared with the observations 
 at 2$\sigma$ level~\cite{WMAP7}
\bea 
 \Omega_{DM} h^2 = 0.1120 \pm 0.0056.  
\label{relicB}
\eea 
Fig.~\ref{fig3} shows the relic abundance of the right-handed 
 neutrino dark matter as a function of its mass. 
The dashed lines correspond to the upper and the lower bounds 
 on the dark matter relic abundance in Eq.~(\ref{relicB}). 
We find two solutions  
\bea 
 M \simeq 906, \; \; 1016 \; {\rm GeV}.   
\eea 
It turns out from Fig.~\ref{fig3} that 
 in order to reproduce the observed relic abundance, 
 the enhancement of the annihilation cross section is necessary, 
 so that the mass of the dark matter should be close 
 to the Z' boson resonance point. 
The dark matter mass $M=906$ GeV coincides 
 with the value presented in the previous section. 
For a different parameter choice, the $Z_2$-odd right-handed 
 sneutrino (scalar or pseudo-scalar) can be the lightest 
 $Z_2$-odd particle and a candidate for the dark matter, 
 instead of the right-handed neutrino. 
However, in this case, there is no $s$-channel Z' boson 
 mediated process and the resultant relic abundance 
 is found to be too large.

\section{Phenomenological constraints}
The $B-L$ symmetry works to forbid phenomenologically dangerous 
 R-parity violating terms in the MSSM. 
Although this is a remarkable advantage of the $B-L$ extended model, 
 the $B-L$ symmetry is eventually broken at the TeV scale, 
 in particular, R-parity violation also occurs 
 by non-zero VEV of the right-handed sneutrino $\tilde{N}^c_3$. 
We here consider phenomenological constraints 
 on our model associated with the R-parity violation.

Once $\langle \tilde{N}^c_3 \rangle \neq 0$ has developed  
 the so-called bilinear R-party violating term 
 is generated in Eq.~(\ref{WBL}), 
\bea 
  W_{BL} \supset 
     \sum_{j=1}^3 y_D^{3j} \langle {\tilde N}^c_3 \rangle L_j H_u. 
\eea 
These terms induce the lepton number violating Yukawa couplings  
  in the MSSM superpotential, 
\bea 
 W_{\Delta L=1} \sim  \left(y_e^{ij} \epsilon^k \right) E^c_i L_j L_k  
                    + \left(y_d^{ij} \epsilon^k \right) D^c_i Q_j L_k, 
\eea 
where $y_e$ and $y_d$ are Yukawa matrices for the charged leptons 
 and the down-type quarks, and 
\bea 
  \epsilon^k = y_D^{3 k} \frac{\langle \tilde{N}^c_3 \rangle}{\mu}
\eea  
 is the mixing parameters between $H_d$ and $L_i$ 
 with the $\mu$-parameter for the Higgs doublets in the MSSM. 
Through the interactions, lepton-number-changing processes 
 can be active in the early universe and erase 
 an existing baryon asymmetry. 
The requirement that this erasure should not occur 
 before the electroweak transition typically 
 gives~\cite{non-erasure}
\bea 
 \left(y_e^{ij} \epsilon^k \right), \; 
 \left(y_d^{ij} \epsilon^k \right) \lesssim 10^{-7}, 
\eea 
 which in turn implies 
\bea 
  \epsilon^k \lesssim 10^{-5} \cos \beta.  
\eea 
This condition is marginally satisfied with 
 the natural order of magnitude for $y_D \sim 10^{-6}$ 
 in the seesaw mechanism at the TeV scale. 
Note that this is a sufficient  condition, and 
 some flavor  structures can relax it~\cite{non-erasure}.

It would be fair to mention that 
 although the renormalizable R-parity violating terms 
 in the MSSM are forbidden by the $B-L$ symmetry, 
 one may introduce dimension five $B-L$ symmetric operators 
 such as $\kappa_{ijk \ell} Q_i Q_j Q_k L_\ell/M_P$. 
Even though these are Planck-scale suppressed operators, 
 they can lead to rapid proton decay. 
In fact, experimental limits from nucleon stability are very severe, 
 $\kappa_{1121}, \kappa_{1122} \lesssim 10^{-8}$~\cite{dim5}. 
We need to assume such small parameters. 
Since the above dimension five operator is also R-parity symmetric, 
 the same discussion is applied to the MSSM.

\section{Conclusions and discussions} 

The minimal gauged U(1)$_{B-L}$ model based on 
 the gauge group 
 SU(3)$_c \times$SU(2)$_L \times$U(1)$_Y \times$U(1)$_{B-L}$  
 is an elegant and simple extension of the Standard Model, 
 in which the right-handed neutrinos of three generations 
 are necessarily introduced for the gauge and gravitational 
 anomaly cancellations. 
The mass of right-handed neutrinos arises associated 
 with the U(1)$_{B-L}$ gauge symmetry breaking, 
 and the seesaw mechanism is naturally implemented. 
Supersymmetric extension of the minimal $B-L$ model offers 
 not only a solution to the gauge hierarchy problem 
 but also a natural mechanism of breaking 
 the $B-L$ symmetry at the TeV scale, namely, 
 the radiative $B-L$ symmetry breaking. 
Although the radiative symmetry breaking at the TeV scale 
 is a remarkable feature of the model, R-parity is also 
 broken by non-zero VEV of a right-handed sneutrino. 
Therefore, the neutralino, which is the conventional 
 dark matter candidate in SUSY models, becomes unstable 
 and cannot play the role of the dark matter any more.

We have proposed introducing a $Z_2$-parity and 
 assigned an odd-parity to one right-handed neutrino. 
This parity ensures the stability of the right-handed neutrino 
 and hence the right-handed neutrino can be a dark matter candidate 
 even in the presence of R-parity violation. 
No new particle introduced for the dark matter is required. 
We have shown that for a parameter set, 
 the mass squared of a right-handed sneutrino 
 is driven to be negative by the RGE running. 
Analyzing the scalar potential with RGE solutions 
 of soft SUSY breaking parameters, we have identified 
 the vacuum where the $B-L$ symmetry as well as R-parity 
 is broken at the TeV scale.

We have numerically integrate the Boltzmann equation for 
 the $Z_2$-odd right-handed neutrino and 
 found that its relic abundance is consistent with the observations. 
In reproducing the observed dark matter relic density, 
 an enhancement of the annihilation cross section 
 via the Z' boson $s$-channel resonance is necessary, 
 so that the dark matter mass should be close 
 to half of Z' boson mass.

Associated with the $B-L$ symmetry breaking, 
 all new particles have TeV-scale masses, 
 which is being tested at the LHC in operation. 
Discovery of the Z' boson resonance at the LHC~\cite{LHCZ'}
 is the first step to confirm our model. 
Once the Z' boson mass is measured, the dark matter mass 
 is also determined in our model. 
If kinematically allowed, the Z' boson decays to 
 the dark matter particles with the branching ratio 
 $\sim 0.6$ \% (see Eq.~(\ref{width})). 
Precise measurement of the invisible decay width of Z' boson 
 can reveal the existence of the dark matter particle.

A variety of experiments are underway to directly or 
 indirectly detect dark matter particles. 
For the detection, it is crucial for a dark matter particle 
 to have sizable spin-independent and/or spin-dependent 
 elastic scattering cross sections with nuclei. 
In our model, the right-handed neutrino dark matter 
 couples with quarks in two ways. 
One is via the Z' boson exchange process, 
 the other is MSSM Higgs boson mediated processes. 
Because of its Majorana  nature, 
 the dark matter particle has the axial vector coupling 
 with the Z' boson, while the SM fermions have the vector couplings. 
As a result, the Z' boson exchange process has no contribution 
 to the elastic scattering between the dark matter particle 
 and quarks in the non-relativistic limit. 
Although the right-handed neutrino dark matter has no direct coupling 
 with the MSSM Higgs bosons, such couplings are generated 
 after the $B-L$ and electroweak symmetry breakings. 
For example, a mixing mass, 
 $|y_3|^2 \langle \Phi \rangle \langle \tilde{N}^c_3 \rangle$, 
 between $\Phi$ and $\tilde{N}^c_3$ 
 is generated by the radiative $B-L$ symmetry breaking 
 and after the electroweak symmetry breaking, 
 another mixing mass such as 
 $|y_D|^2  \langle \tilde{N}^c_3 \rangle \langle H_u \rangle$ 
 between $\tilde{N}^c_3$ and $H_u$ is generated. 
These mixing masses eventually induce 
 the couplings between the right-handed neutrino dark matter 
 and the MSSM Higgs bosons. 
However, the coupling constants are suppressed 
 by the small factor $|y_D|^2 \sim 10^{-12}$, 
 so that the spin-independent cross section 
 for the elastic scattering off nuclei 
 mediated by the MSSM Higgs boson exchanges 
 is too small to be detected.

\section*{Acknowledgments}
The work of N.O. is supported in part 
 by the DOE Grants, No. DE-FG02-10ER41714.


\end{document}